\documentclass[english,aps,prd,showpacs,twocolumn,superscriptaddress,preprintnumbers,nofootinbib]{revtex4-1}
\global\long\def\order#1{\mathcal{O}\left(#1\right)}
\global\long\def\d{\mathrm{d}}

\def\GLL{\Gamma_{\text{CL}}}
\def\NLL{N_{\text{CL}}}
\def\GVP{\Gamma_{\text{VP}}}
\def\NVP{N_{\text{VP}}}
\def\GLO{\Gamma_{\text{LO}}}
\def\NLO{N_{\text{LO}}}
\def\dVP{\delta_{\text{VP}}}
\def \BVP{B_{\text{VP}}}

\def\Emax{E_{\text{max}}}
\def\Zal{Z\alpha}

\usepackage{amsmath}
\usepackage{amssymb}

\usepackage{slashed}
\usepackage{graphicx}
\usepackage{epsfig}

\usepackage{epstopdf}
\usepackage{esint}

\begin{document}

\title{Bound muon decay spectrum in the leading logarithmic accuracy}

\preprint{Alberta Thy 7-16}
\author{Robert Szafron}
\email{szafron@ualberta.ca} 
\author{Andrzej Czarnecki}
\affiliation{Department of Physics, University of Alberta, Edmonton,
  Alberta, Canada T6G 2G7}

\begin{abstract}
  We compute the dominant, logarithmically enhanced radiative
  corrections to the electron spectrum in the bound muon decay in the
  whole experimentally interesting range. The corrected spectrum fits
  well the TWIST results. The remaining theoretical error, dominated
  by the nuclear charge distribution, can be reduced in the
  muon-electron conversion searches by measuring the spectrum slightly
  below the New Physics signal window.
\end{abstract}
\maketitle

\section{Introduction}

The spectrum of electrons from the decay of a muon bound in an
atom  (decay in orbit, DIO) has two parts. The low-energy part,
$E_e \lesssim m_\mu/2$ ($E_e$ is the electron energy and $m_\mu$ is
the muon mass) is present also in the free muon decay but is
reshuffled because the bound muon is moving and the daughter electron
interacts with the electric field of the nucleus. This modification of
the spectrum was observed by TWIST \cite{Grossheim:2009aa}. 

In addition, the possibility of transferring momentum to the nucleus
approximately doubles the range of energy accessible to the electron,
adding the high-energy region $m_\mu/2 \lesssim E_e \lesssim m_\mu$.
This region has already been explored for some nuclei including
titanium, sulfur and gold  \cite{Bertl:2006up,
  Dohmen:1993mp,Ahmad:1988ur,Burnham:1987gr, Badertscher:1981ay}.
Upcoming experiments COMET and Mu2e \cite{Bartoszek:2014mya,Cui:2009zz,Kuno:2013mha} will  measure it for aluminum with
a high precision.   Their main goal is
to discover the lepton-flavor violating muon-electron
conversion. High-energy electrons from the DIO are a background in this search.

The purpose of this paper is to improve the theoretical description of
the entire spectrum by determining logarithmically enhanced radiative
corrections. We focus on aluminum, the stopping
material in COMET and Mu2e.

The spectrum of the free muon decay is known including corrections of
the first  \cite{Behrends:1956mb} and  of the second
order \cite{Anastasiou:2005pn} in the fine structure constant
$\alpha\approx {1}/{137}$,
as well as the leading logarithms in the third order $\left(\frac{\alpha}{\pi}\ln\frac{m_{\mu}}{m_{e}}\right)^{3}$ \cite{Arbuzov:2002rp}.
In the case of the bound muon, even $\order{\alpha}$
effects are known only in a limited range of electron energies. 
Radiative corrections near the top  of the electron spectrum in
the free muon decay ($E_{e}\sim\frac{m_{\mu}}{2}$) were evaluated
in \cite{Czarnecki:2014cxa} by convoluting the $\order{\alpha}$
free-muon spectrum with the so-called  shape function
\cite{Szafron:2015mxa}, reconciling TWIST results
\cite{Grossheim:2009aa} with quantum electrodynamics.  This approach
had been developed in the heavy quark effective theory  to describe
decays of $B$-mesons \cite{Neubert:1993ch,Bigi:1993ex} using factorization theorems.
Alas, the factorization cannot be applied
when the electron energy is much larger than the half of the muon
mass. 

Fortunately, the highest energy endpoint of
the DIO spectrum ($E_{e}\sim
m_{\mu}$) offers a different simplification: one can expand in
the number of photons exchanged with the nucleus, parameterized by
$Z\alpha$ where $Z$ is the proton number of the nucleus (for aluminum, $Z=13$).
Radiative corrections in the endpoint region have been evaluated
in \cite{Szafron:2015kja}. It is still unknown how to compute full
radiative corrections for intermediate electron energies $\frac{1}{2}m_{\mu}<E_{e}<m_{\mu}$
 \cite{Szafron:2015wbm}.

However, the likely largest corrections can be computed.
Enhanced effects
$\sim\frac{\alpha}{\pi}\ln\frac{m_{\mu}}{m_{e}}$ arise from
collinear photons and can be found from collinear
factorization theorems \cite{Ellis:1991qj}, without a
new loop calculation. 

In a muonic atom, vacuum polarization (VP) is an additional source of
large logarithms. It modifies the Coulomb potential and is taken into
account numerically together with the effects of the finite charge
distribution in the nucleus. Typically, the VP correction is on the
order of $\frac{\alpha}{\pi} \ln\frac{Z\alpha
  m_{\mu}}{m_{e}}$, or
$\frac{\alpha}{\pi} \ln\frac{m_{\mu}}{m_{e}}$  near the DIO endpoint.

Section \ref{sec:2} presents details of both collinear and VP
corrections. Section \ref{sec:3} summarizes numerical results. In Section \ref{sec:4}
we discuss the uncertainty due to the nuclear charge distribution
and suggest a means of lowering it.

\section{Leading logarithmic corrections to the DIO spectrum}
\label{sec:2} 
We assume that the daughter electron is relativistic.  For a
low-energy electron, additional non-perturbative phenomena would have
to be considered. For example, a slow electron can be captured into
the atom in the final state.  The low-energy part of the spectrum is
not yet fully understood \cite{Greub:1994fp}. However, it involves only
a small fraction of electrons because of the phase space suppression
and it is not relevant for conversion experiments, sensitive only to the
high-energy region of the spectrum.

\subsection{Collinear photons}
Focusing on an energetic electron, we first consider the emission of
collinear photons. For electron energies much larger than
the electron mass $E_{e}\gg m_{e}$, Coulomb corrections are small
and do not affect the collinear limit of the amplitude in the leading
order in $Z\alpha$. Before
the emission of a collinear photon, the electron is almost on-shell
and propagates over distances large compared with the size of the muonic
atom. Hence, the collinear emission is a long-distance phenomenon that takes
place after the electron escapes the region of the
strong binding potential. On the other hand, if the photon
is emitted before the last scattering of an electron on the nucleus,
the electron is still off-shell and the amplitude is not
singular. In such case, the corrections are not enhanced by a
large logarithm $\ln\frac{m_{\mu}}{m_{e}}$ or are suppressed
by additional powers of $\Zal$. For aluminum,
$\Zal \ln\frac{m_{\mu}}{m_{e}}\sim\frac{1}{2}$.  These corrections
are comparable with the non-logarithmic term
$\order{\alpha }$ and we neglect them.

Collinear corrections can be calculated using a
factorization theorem, previously employed to improve
 the free muon spectrum \cite{Arbuzov:2002pp,Arbuzov:2002cn,Arbuzov:2002rp}.
Following Ref.~\cite{Arbuzov:2002pp} we evaluate the 
collinear logarithms $\frac{\d \GLL}{\d E_{e}}$
convoluting the leading order spectrum $\frac{\d \GLO}{\d E_{e}}$
with the electron structure function,
\begin{equation}
\frac{\d \GLL}{\d E_{e}}=\frac{\d \GLO}{\d E_{e}} \otimes D_{e} +\order{\Zal\frac{\alpha}{\pi}\ln\frac{m_{\mu}}{m_{e}}},\label{eq:conv}
\end{equation}
where 
\begin{equation}
D_{e}(x)=\delta(1-x)+\frac{\alpha}{2\pi}\left(\ln\frac{m_{\mu}^{2}}{m_{e}^{2}}-1\right)P_{e}\left(x\right)+\order{\alpha^{2}}
\end{equation}
and the electron splitting function is 
\[
P_{e}(x)=\left[\frac{1+x^{2}}{1-x}\right]_{+}.
\]
We employ the dimensionless variable
$x=E_{e}/\Emax$ with $\Emax$ denoting
the maximum electron energy. The convolution is defined as $A\otimes
B\left(z\right)=\int_{0}^{1}\d x\int_{0}^{1}\d y\delta\left(z-xy\right)A(x)B(y)$
and the leading order term $\frac{\d \GLO }{\d E_{e}}$ includes
Coulomb effects to all orders in $\Zal$. 

Eq.~(\ref{eq:conv}) ensures a cancellation of the mass singularity
in the correction to the bound muon lifetime, in agreement with the
Kinoshita-Lee-Nauenberg theorem \cite{Kinoshita:1962ur,Lee:1964is}.

\subsection{Vacuum polarization}

The second type of large logarithms comes from the vacuum
polarization that strengthens the binding. The VP does
not contribute to the free muon spectrum at one loop, hence this
correction is exclusive to the bound muon and related to
other binding effects. The $\order{\alpha}$ correction to the
potential is known as the Uehling term \cite{Uehling:1935uj}.  For a
pointlike nucleus the binding potential is
\begin{widetext}
\begin{equation}
V(r)=-\frac{\Zal}{r}\left(1+\frac{2\alpha}{3\pi}\int_{1}^{\infty}\d
  xe^{-2m_{e}rx}\frac{2x^{2}+1}{2x^{4}}\sqrt{x^{2}-1}\right)+\order{\left(\Zal\right)^{3}}.
\end{equation}
\end{widetext}
 The VP potential for an arbitrary charge distribution
is given in Ref.~\cite{Hylton:1985zz}. 
The electron loop modifies the potential at distances comparable
to the Compton wavelength of electron $r_{e}\sim\frac{1}{m_{e}}$
or smaller. The VP term
reduces the endpoint energy and increases the number of high-energy
electrons. It also shrinks the muon orbit. As a result, the muon kinetic
energy and the lifetime increase. 

In muonic atoms, the VP effect   is much larger than
in ordinary atoms \cite{Borie:1982ax}. The binding
energy,
\begin{equation}
E_{b}\simeq -m_{\mu}\frac{\left(\Zal\right)^{2}}{2},
\label{eq:Eb}
\end{equation}
receives a correction that is not suppressed by extra powers of $\Zal$. In
an ordinary atom the Lamb shift contributes at the $(\Zal)^4$
order. For example, the VP correction   starts with
$-\frac{4}{15}\frac{\alpha}{\pi}\left(\Zal\right)^{4}m_{e}$.
This behaviour follows from the range of the VP potential, much
smaller than the size of the electron orbit
$\frac{1}{m_{e}}\ll\frac{1}{m_{e}\Zal}.$ When the electron is replaced
by a muon, the potential reaches beyond the muon orbit,
$\frac{1}{m_{e}}\gg\frac{1}{m_{\mu}\Zal}$, and the correction to the
binding energy behaves as
$\sim\frac{\alpha}{\pi}(\Zal)^2\ln\frac{m_{\mu}\Zal}{m_{e}}$.  For a
pointlike nucleus and using non-relativistic muon wave function, in
the limit $m_{e}\ll m_{\mu}$, we get the correction to the binding energy
\begin{eqnarray}
\Delta \BVP &=&\frac{\alpha}{\pi}\left(\Zal\right)^{2}m_{\mu}
\left(\frac{11}{9}-\frac{2}{3}\ln\frac{2m_{\mu}\Zal}{m_{e}}\right)
  \nonumber \\
&=&-2.7 \text{ keV for aluminum.}
\label{eq:DeltaE}
\end{eqnarray}
This is larger than the total uncertainty in the binding energy and
has to be included in the evaluation of the endpoint energy. We will
return to this correction in the discussion of numerical results in
Section \ref{sec:3}.

The logarithmic terms in the VP correction can also be reproduced
by using in eq.~(\ref{eq:Eb}) the running coupling constant
$\alpha\left(Q^{2}\right)=
\frac{\alpha}{1-\frac{\alpha}{3\pi}\ln\left(\frac{Q^{2}}{m_{e}^{2}}\right)}$,
with $\sqrt{Q^2}=m_{\mu}\Zal$.

In the DIO spectrum, the VP effects do not factorize, unlike the
collinear corrections. They are accounted for, together with the
finite nuclear size, by numerically solving the Dirac equation.

Large corrections with logarithms of $\Zal$ are also present
in the DIO spectrum. Pure relativistic corrections can contain $\ln \Zal$,
typically suppressed by two powers of $\Zal$ \cite{Szafron:2013wja}.
In our numerical approach, we solve the Dirac equation without any non-relativistic
expansion; hence, these terms are automatically included in our leading
order spectrum.

Logarithms of $\Zal$ will appear also in radiative corrections
involving ultra-soft photons, like in the classical calculation of the
Lamb shift \cite{Pineda:1997ie}. These Bethe-type logarithms are
suppressed by additional powers of $\Zal$, in the same manner
as the VP shift of the binding energy in electron atoms. 

\section{Numerical results}
\label{sec:3}

We compute wave functions of the bound muon and of the daughter
electron by numerically solving the Dirac equation
\cite{rosebook,BetaWF}. An analytical solution is not known for a
realistic distribution of the nuclear charge density.  In order to
find the DIO spectrum we have implemented in Python matrix elements,
including nuclear recoil corrections, given in
\cite{Czarnecki:2011mx}.  As a check of the numerical code, we have
compared the muon binding energy and the DIO spectrum with previous
results \cite{Czarnecki:2011mx,Watanabe:1987su}.

We use the Fermi model for the nuclear charge density
distribution fitted to the electron elastic scattering data
\cite{vries87}. In Section \ref{sec:4} we discuss other possible
models and the uncertainty related to the nuclear charge radius. 

Near the endpoint, the spectrum rapidly varies with
energy, like $(\Emax-E_e)^5$, so a precise value of $\Emax$ is
critical. 
We predict, including the charge distribution corrections, the VP
term, and the recoil correction (see eq.~(13) in \cite{Czarnecki:2011mx}),
\begin{equation}
\Emax=m_{\mu}-E_\text{rec}+E_{b}=104.971(1)\;\text{MeV}.
\label{eq:Emax}
\end{equation}
The error comes from the uncertainty in the charge distribution. 
The difference between our result and the endpoint energy without the VP
correction, $E_{\mu e}=104.973(1)$ \cite{Czarnecki:2011mx},
is consistent with Eq.~(\ref{eq:DeltaE}). 
\begin{figure}
\noindent \begin{centering}
\includegraphics[width=0.5\textwidth]{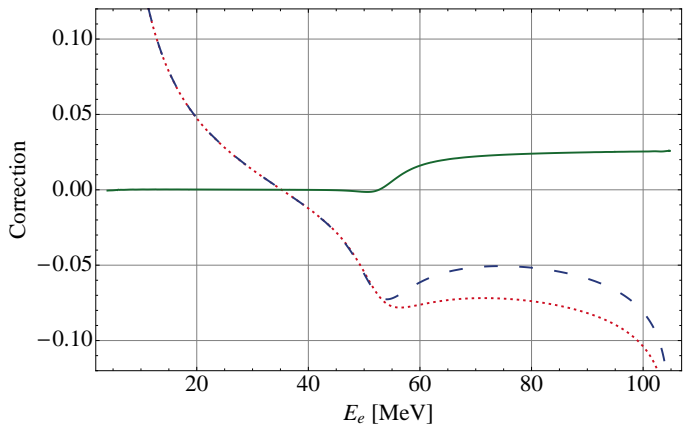}
\par\end{centering}
\caption{\label{fig:LLcor}Leading corrections to the bound muon
  spectrum. The VP correction $\frac{\NVP-\NLO}{\NLO}$ is
  mostly positive (solid line). We have shifted the electron energies
  in the VP term,
  $E_e\rightarrow E_e+\Delta \BVP$, so that the $\NVP$ and $\NLO$
  spectra have a common endpoint energy, $E_{\mu e}$.  The correction
  $\frac{\NLL-\NLO}{\NLO}$, due to collinear photons, decreases the
  number of electrons near the endpoint (dotted line). The dashed line
  represents the total correction. See text for details.  }
\end{figure}

We denote $\NLL =\frac{\d \GLL}{\d E_{e}}$ and
$\NLO=\frac{\d \GLO }{\d E_{e}}$. We also introduce
$\NVP =\frac{\d \GVP}{\d E_{e}}$ as the leading order spectrum that
includes the VP correction only. Relative collinear 
$\frac{\NLL-\NLO}{\NLO}$ and VP 
$\frac{\NVP-\NLO}{\NLO}$ corrections are presented in Fig.~\ref{fig:LLcor}. For a
1~MeV signal window near the endpoint, we obtain a 12\% reduction of
the number of DIO events, consistent with \cite{Szafron:2015kja}. 

Should the DIO spectrum be needed for electron energies closer to the
endpoint, we recommend using eq.~(2) in
\cite{Szafron:2015kja}, where soft photons have been exponentiated.

For low electron energies, $E_e \leq \frac{m_{\mu}}{2}-m_\mu \Zal$,
the DIO spectrum is dominated by the free muon decay corrected by the
binding effects.  Consequently, in this region, the VP is a relatively
unimportant subleading effect.  The VP correction becomes significant
for electron energies above the free muon endpoint,
$E_{e}>\frac{m_{\mu}}{2}$, where the free muon spectrum is
absent. Here, the spectrum is dominated by the binding effects and is
sensitive to the details of the binding potential. The VP correction
is particularly important near the endpoint $E_e \simeq \Emax$ where
highly-virtual Coulomb photons transfer a large momentum to the
nucleus.

\begin{figure}
\begin{centering}
\includegraphics[width=0.99\columnwidth]{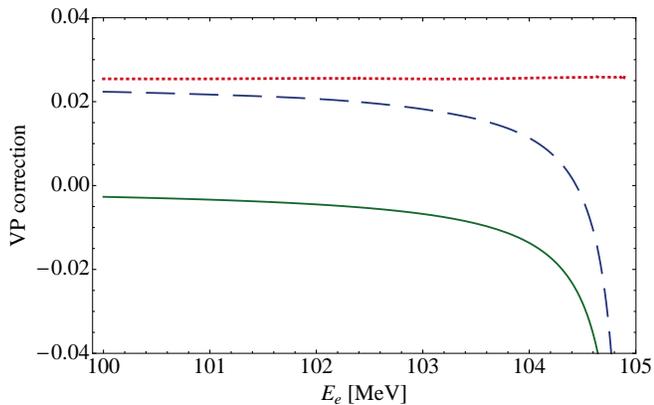}

\caption{\label{fig:VPend} Vacuum polarization correction to the DIO
  spectrum around the endpoint. The solid line illustrates the relative
  change of the spectrum due to the decrease of the endpoint energy,
 $\left(\frac{\Emax-E_e}{E_{\mu e}-E_e}\right)^5-1$. The dotted line
  is a correction due to the shift in the wave functions $\dVP$, see
  Eq.~(\ref{eq:DVP}). The dashed line shows both effects combined. }

\end{centering}

\end{figure}

In Fig.~\ref{fig:VPend} we show the VP correction in the DIO endpoint
region, where the spectrum has a simple dependence on the electron
energy,
\begin{equation}
\frac{\d \Gamma}{\d E_{e}}\equiv N\sim\left(\Emax-E_{e}\right)^{5}.
\label{eq:delta^5}
\end{equation}
Here the VP correction manifests itself in two ways.
The correction due to the shift in the endpoint energy (\ref{eq:DeltaE})
is important only very close to $\Emax$. For smaller electron
energies it becomes negligible.  To illustrate this effect we plotted
$\left(\frac{\Emax-E_e}{E_{\mu e}-E_e}\right)^5-1$ in
Fig.~\ref{fig:VPend}. 

The VP also modifies muon and electron wave functions. This effect
does not depend strongly on the electron energy, because it
is dominated by the running of the coupling constant
$\alpha\left(Q^{2}\right)$. It varies slowly for
$\left|Q^{2}\right|\sim m_{\mu}^{2}$.  

To quantify this effect, we define
the shift of the spectrum due to the VP correction to the wave functions as 
\begin{equation}
\dVP=\frac{\NVP \left(E_{e}+\Delta \BVP\right)-\NLO
  \left(E_{e}\right)}{\NLO \left(E_{e}\right)}\label{eq:DVP}
\end{equation}
in the limit when $E_{e}$ approaches the endpoint of the leading order
spectrum $E_{\mu e}$. We find
numerically $\dVP=2.5\%$. Ref.~\cite{Szafron:2015kja} provided the correction to
the leading term of the spectrum expanded in $\Zal$:
$\delta_{\text{VP},\left(\Zal\right)^{5}}=2.9\%$.  The difference with
$\dVP$ is caused by higher orders in $\Zal$, estimated as minus 20\%
of the leading
$\left(\Zal\right)^{5}$ term \cite{Szafron:2015kja}.

\begin{figure*}[t]
\begin{centering}
 \includegraphics[height=0.6\columnwidth]{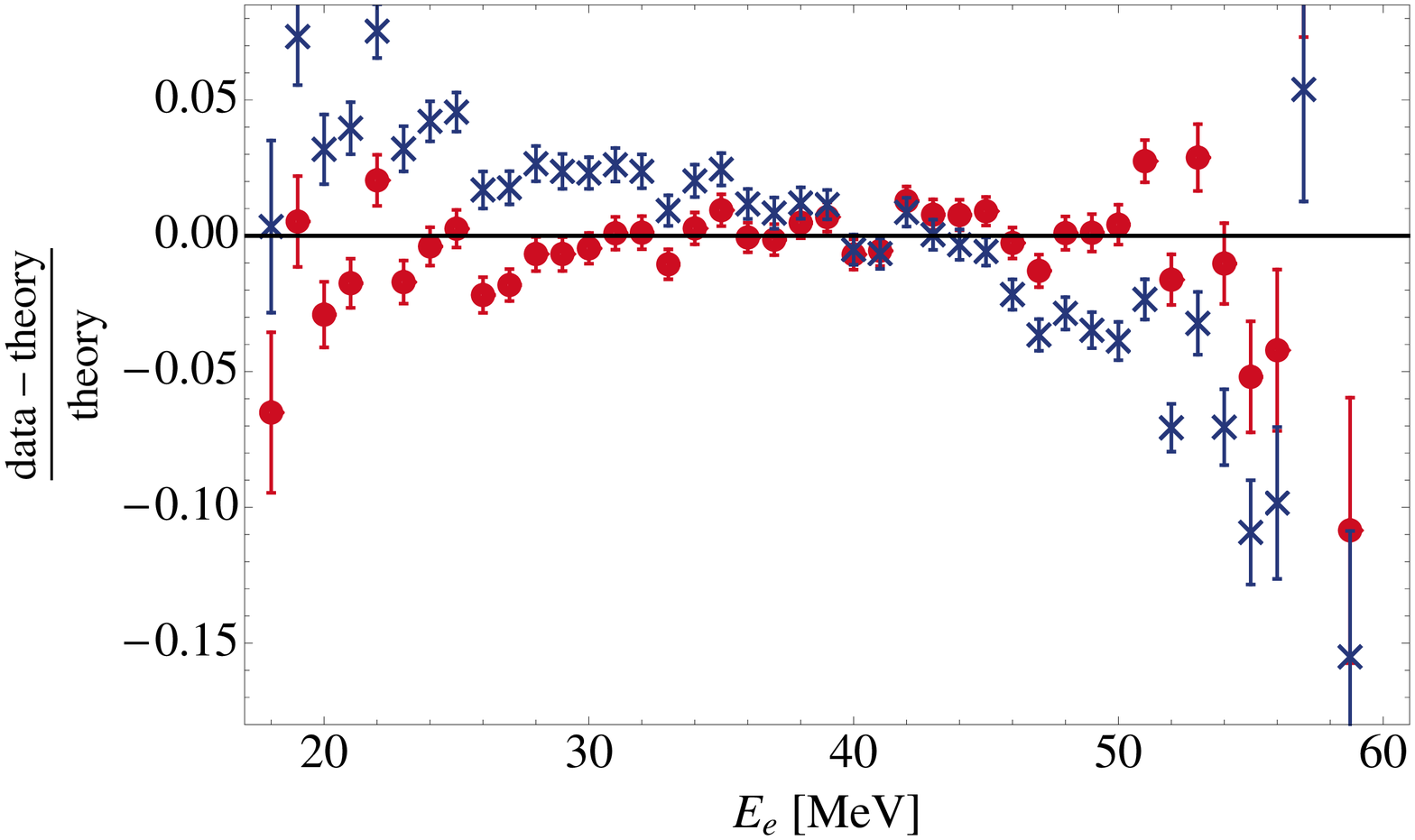} \includegraphics[height=0.6\columnwidth]{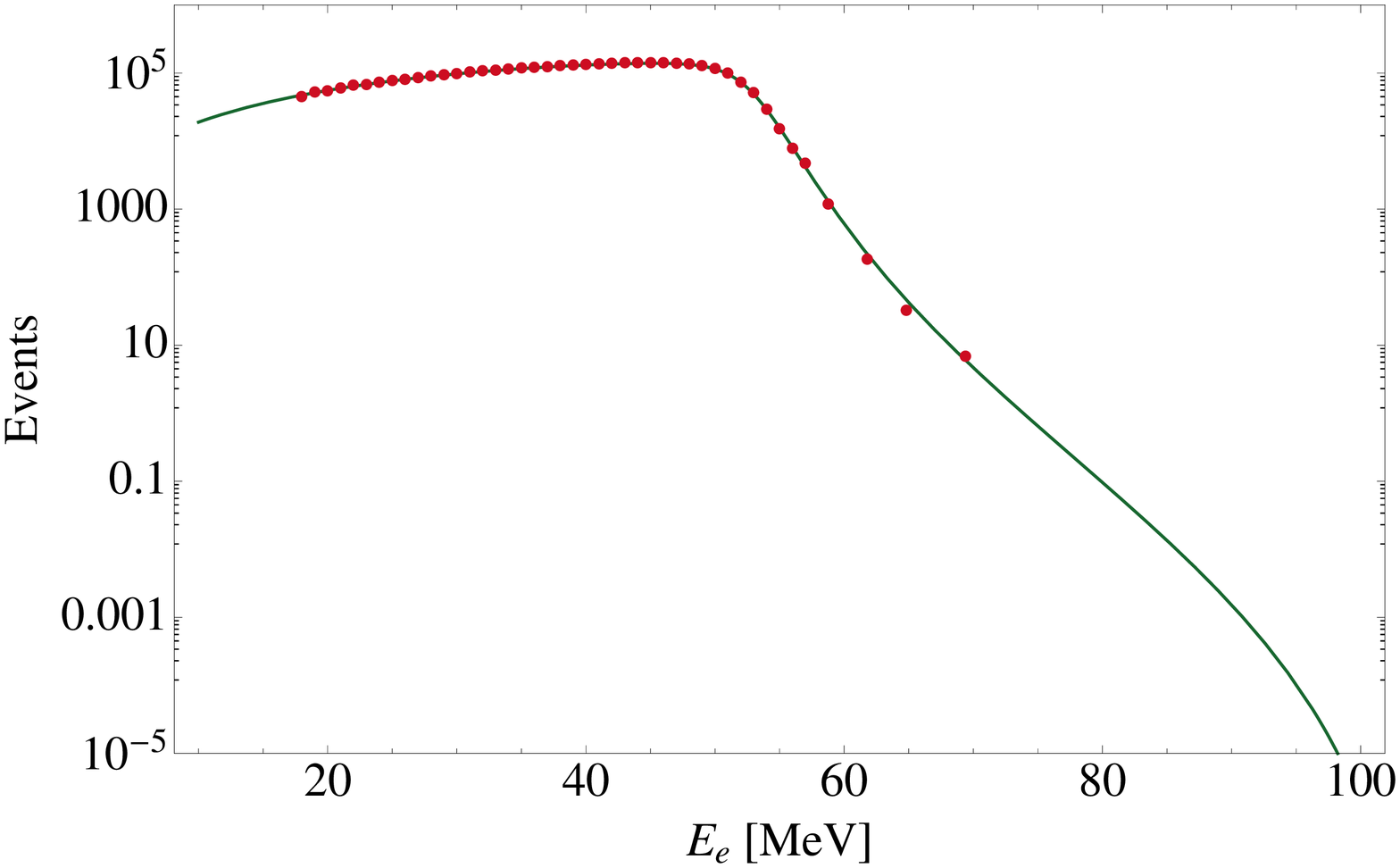}
\end{centering}

\caption{\label{fig:TWIST}Left panel: the difference between TWIST data
and the theoretically calculated spectrum normalized to our theoretical
evaluation of the DIO spectrum. Crosses represent the leading order evaluation
without any radiative corrections. Dots correspond to our new evaluation
that includes leading logarithmic corrections. Right panel: the
DIO spectrum (solid line) fitted to TWIST data (dots).  With the
results of this paper, a measurement of
the DIO spectrum at  energies
$E_e\sim \frac{m_\mu}{2}$ can be used to calibrate the energy response in the future conversion experiments. }
\end{figure*}

As a check, we have compared our results with the spectrum measured by
TWIST \cite{Grossheim:2009aa}. The results are presented in
Fig.~\ref{fig:TWIST}. The quality of the fit is comparable to our
previous approach based on the shape function (see Fig. 3 in
\cite{Czarnecki:2014cxa}).  However, now we are not limited to
electron energies $E_{e}\lesssim\frac{m_{\mu}}{2}$. Also, we are not
including the energy scale uncertainty in the fit. This reduces the
number of fit parameters to one: the overall normalization of the
spectrum.  The quality of the fit is characterized by $\chi^2$ per
degree of freedom, $\chi^2$/dof. The leading order DIO spectrum gives
$\chi^2$/dof=8.8.  When radiative corrections are included using the
shape function, this decreases to $\chi^2$/dof=3.9.  The spectrum
obtained in the present paper gives a slightly better
$\chi^2$/dof=2.8. The quality of the fit could likely be improved by
including the TWIST systematic errors and correlations among energy
bins but we are not qualified to do this.

\section{Nuclear charge distribution}
\label{sec:4}
In order to quantify the uncertainty due to the nuclear charge
density, we have examined three experimental sources. Two use the Fermi model, 
\begin{equation}
\varrho(r)=\frac{\varrho_{0}}{1+\exp\left(\frac{r-r_{0}}{a}\right)},\label{eq:Fermi}
\end{equation}
where $r_0$ is a fitted parameter describing the radius of the
distribution and $a$ is related to the so-called skin thickness.
Elastic electron scattering gives \cite{vries87} $r_{0}=2.84(5)$~fm
for $a= 0.569$~fm and transitions in muonic aluminum
\cite{Fricke:1992zza} give, more precisely, $r_{0}=3.0534(13)$~fm for
$a= 0.523$~fm. Even though the radii seem to differ by more than
one standard deviation, they have been fitted at different values of
the parameter $a$. In our calculation these differences partially
compensate one another, as we shall see below.    

Another  parameterization employs the  spherical Bessel function $j_{0}$,
\begin{equation}
\varrho(r)=\sum_{n}a_{n}j_{0}\left(\frac{n\pi r}{R}\right),\;r<R;\;\;\varrho(r)=0,\;r\geq R,\label{eq:Bessel}
\end{equation}
where $R$ is a cutoff beyond which the density is assumed to be zero,
taken to be $R=7$~fm, and the  coefficients $a_{i}$, for $i=1,\dots,12$,
are given in \cite{vries87}. Unfortunately, no error estimate seems to
be available for $a_i$. 


\begin{figure}[ht]
\noindent \begin{centering}
 \includegraphics[width=0.99\columnwidth]{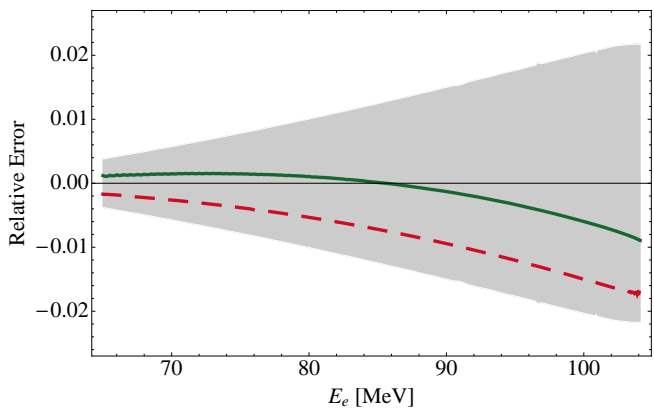}
\par\end{centering}

\caption{\label{fig:Error} Relative uncertainty in the DIO spectrum near the endpoint
related to nuclear charge distribution. The shaded region is obtained
by varying $r_{0}$ in (\ref{eq:Fermi}) within limits
obtained from electron scattering data \cite{vries87}. Solid line
corresponds to the Bessel parametrization (\ref{eq:Bessel}) and the dashed
line to the Fermi distribution obtained from muonic
atoms \cite{Fricke:1992zza}.}
\end{figure}

Only the high-energy part of the  DIO spectrum is sensitive to the
smearing of the nuclear charge. Fig.~\ref{fig:Error} shows the
predictions of the three models in that region, consistent within the
electron scattering errors. 

Electron scattering data give the largest error for the
charge density. 
We use them to tabulate the
DIO spectrum for aluminum including leading logarithmic
corrections; see supplemental material \cite{DIO}. This choice is further justified  by
the muon DIO amplitude near the endpoint being proportional to the
elastic scattering amplitude \cite{Szafron:2015kja}.  To quantify the
dependence
of the error on the electron energy, we approximate the one-sigma boundaries
(the shaded region in Fig.~\ref{fig:Error}) by 
\begin{equation}
\frac{\Delta N}{N} \approx \sigma \frac{2E_{e}-\Emax}{\Emax},\;\text{with}\;\sigma=0.022.\label{eq:DNLL}
\end{equation}
The coefficient 2 in front of $E_e$ reflects the approximate vanishing
of the sensitivity to the nuclear distribution in the low-energy
region: at $E_e=\Emax/2$ and below.

The uncertainty (\ref{eq:DNLL}) can be reduced by measuring, in
conversion experiments, the DIO spectrum outside the conversion signal
window.  To fit $r_0$, such measurements should use the radiatively
corrected DIO spectrum; not only are the corrections large but they also
change the simple functional form of the DIO spectrum near the
endpoint, Eq.~(\ref{eq:delta^5}).

To achieve the necessary accuracy, the DIO spectrum measurement
requires a precise energy calibration. The endpoint energy has
been calculated with a precision of $1$~keV, see Eq.~(\ref{eq:Emax}).
The upcoming experiments will measure the endpoint energy with a larger uncertainty,
$\Delta \Emax$. From Eq.~(\ref{eq:delta^5}) we estimate
how $\Delta \Emax$ influences the number of electrons with
energy $E_{e}=\Emax-\delta$. Denoting the uncertainty in the spectrum
due to $\Delta \Emax$ as $\Delta N_{E}$ we get
\begin{equation}
\frac{\Delta N_{E}}{N}\approx5\frac{\Delta \Emax}{\delta},\label{eq:DNE}
\end{equation}
hence in order to constrain the error related to the nuclear charge
distribution, the number of DIO events $N$ with energy $E_{e}$ has
to be measured with an experimental precision $\Delta N_{\exp}$ such
that 
\begin{equation}
\sqrt{\left(\frac{\Delta N_{\exp}}{N}\right)^{2}+\left(\frac{\Delta N_{E}}{N}\right)^{2}}\lesssim\frac{\Delta N}{N}
\end{equation}
It is most efficient
to use the DIO measurement at an energy for which the largest 
 $\frac{\Delta N_{\exp}}{N}$ can be tolerated. This optimal energy can be calculated
using (\ref{eq:DNLL}) and (\ref{eq:DNE}),
\begin{eqnarray}
  \label{eq:1}
\frac{E^\text{opt}_{e}}{\Emax}&=&1- \xi - \frac{2}{3}\xi^2   - \frac{4}{3}\xi^3 +\order{\xi^4}  
\nonumber\\
\xi & \equiv &
\sqrt[\leftroot{-1}\uproot{2}\scriptstyle 3]{\frac{25}{2} \left(\frac{\Delta \Emax}{\sigma\Emax}\right)^2}
\end{eqnarray}
For example, for $\Delta \Emax=30\;\text{keV}$, the optimal
energy is 90 MeV, and the experimental uncertainty should be smaller than
$\frac{\Delta N_{\exp}}{N}\lesssim 0.012$.

\section{Conclusions}
\label{sec:conc}

We have calculated the energy spectrum of electrons in the bound muon decay including leading logarithmic
corrections. 
For electron energies $E_{e}>100\;\mbox{MeV}$
the sum of vacuum polarization and collinear photon effects decreases
the number of DIO events by more than 10\%, in agreement with the  
endpoint expansion \cite{Szafron:2015kja}. 

Our present result is valid in the entire energy range $E_{e}\gg
m_{e}$ and can be used to calibrate the energy in conversion experiments.
This was not possible with previous results available only near the
endpoint $\Emax$ and in the low-energy (shape function) region,
without means to interpolate in the remaining high-energy region.

The dominant remaining uncertainty comes from the nuclear charge 
distribution. Here new input from experiments
is required. We suggest that the DIO spectrum be used to constrain
the charge distribution, as a byproduct of the conversion search. 

We have neglected the screening by the electron cloud (see
\cite{Czarnecki:2011mx}  for a discussion). In order
to further improve the theoretical description of the DIO spectrum,
this effect should be included together with non-logarithmic radiative
corrections $\order{\frac{\alpha}{\pi}}$.

\begin{acknowledgments}
This research was supported by Natural Sciences and Engineering
Research Council (NSERC) of Canada.  R.S.~acknowledges  support
by the Fermilab Intensity Frontier
Fellowship. Fermilab is operated by Fermi Research Alliance, LLC under
Contract No.~De-AC02-07CH11359 with the United States Department of
Energy.
\end{acknowledgments}

\end{document}